\documentclass[twocolumn]{article}
\usepackage{epsfig}
\usepackage{amsmath}
\begin{document}

\title{Continuum elastic modeling of graphene resonators}

\author{Juan Atalaya,
  Andreas Isacsson\footnote{Corresponding author: andreas.isacsson@chalmers.se.}, and Jari M. Kinaret\\
  {\it Department of Applied Physics, Chalmers University of Technology}\\{\it SE-412 96 G{\"o}teborg Sweden}.}
\date{\today}
\maketitle

\begin{abstract}
  Starting from an atomistic approach we have derived a hierarchy of
  successively more simplified continuum elasticity descriptions for
  modeling the mechanical properties of suspended graphene sheets. The
  descriptions are validated by applying them to square graphene-based
  resonators with clamped edges and studying numerically their
  mechanical responses.  Both static and dynamic responses are
  treated. We find that already for deflections of the order of 0.5{\AA}
  a theory that correctly accounts for nonlinearities is necessary and
  that for many purposes a set of coupled Duffing-type equations may
  be used to accurately describe the dynamics of graphene membranes.
\end{abstract}

{\bf Introduction.} Recent progresses in fabricating graphene, a
monolayer of graphite, have considerably boosted the attention for
this material \cite{Graphene_Novoselov}. Among its unique features are
remarkable electronic properties \cite{Graphene_Guinea, RPM}, which make
graphene of considerable interest for both fundamental science and
technological applications. Moreover, its exceptionally large
mechanical strength \cite{Graphene_Bunch} and ability to sustain large
electrical currents can be of great value in the broad field of
nanodevices. In particular, in the field of nanoelectromechanical
systems (NEMS), graphene-based mechanical resonators were recently
demonstrated \cite{Graphene_Bunch, McEuen3} and theoretical work indicates a
strong coupling between deformation and intrinsic electronic
properties of graphene \cite{Kim,Isacsson}.

Most research on graphene has hitherto focused on the electronic
properties of graphene, and less attention has been directed to
mechanical properties.  For modeling NEMS a reliable and efficient
description of the mechanical response of nanocarbons to external
forces is essential \cite{CNT_Yuri,CNT_Kinaret}.  While continuum
elasticity theory has been applied succesfully to the study of
mechanical properties of nanotubes for a long time,
\cite{CNT_Arroyo,CNT_Poot,CNT_Wilber} it has only recently been
applied to graphene membranes \cite{Graphene_Huang, Graphene_Cristiano,
 Samhedi, McEuen_vac}.  

In this work we first  formulate a
general nonlinear elasticity theory for graphene sheets starting from
an atomistic description in terms of a valence force field model. 
Through successive approximations, we then derive
simplified models which are easier to solve and to study analytically.
We then apply these continuum descriptions to a drum-like resonator and investigate the
mechanical response in both the static and dynamic cases.
The results obtained
agrees well with the experimentally observed responses \cite{Poot,McEuen2}.

\noindent {\bf Continuum elasticity model for graphene.} The interaction
potential between carbon atoms in graphene can be modeled by a valence
force field model \cite{Graphene_Lobo,Graphene_Keating} where the
potential energy $U_{sp^2}$ between $sp^2$ bonds is given by
\begin{eqnarray} \label{eq:Eq_Usp2} 
&&U_{sp^2} = \frac{1}{2} \sum_{i=1}^{N_{at}} \sum_{j} \frac{\alpha}{4a^2_0} \left(\bar{r}_{ij}^2-a^2_0 \right)^2  \\
&&+ \sum_{i=1}^{N_{at}} \sum_{j < k}  \frac{\beta}{a^2_0} \left(\bar{r}_{ij} \cdot \bar{r}_{ik}+\frac{1}{2} a^2_0 \right)^2  + \sum_{i=1}^{N_{at}} \gamma \bar{D_i} \cdot \bar{D_i}.\nonumber
\end{eqnarray}
Here the index $i$ labels the carbon atoms while indices $j$ and $k$
are bond labels for the nearest neighbor atoms of $i$. Thus,
$\bar{r}_{ij}$, $j=1,2,3$, are the three bond vectors that connect the
atom $i$ to its nearest neighbors. The parameters $\alpha$, $\beta$
and $\gamma$ are constants, $N_{at}$ the number of atoms,
$a_{0}=1.421$ {\AA} is the equilibrium bond length in graphite, and
$\bar{D_i} = \sum_{j=1}^3\bar{r}_{ij}$ is the dangling bond vector.
The first two terms of Eq. \eqref{eq:Eq_Usp2} represent the energy
cost necessary to change the length and angle between covalent C-C
bonds. In the continuum description, we refer to it as the stretching
energy. The last term of Eq. \eqref{eq:Eq_Usp2} gives the energy cost
necessary to change the angle between $p_z$-orbitals, which are
approximately normal to the graphene surface.  We refer to 
energy contributions from this term as bending energy.

Provided that the length scale of the deformation is large compared to
the lattice spacing (long wavelength limit), continuum theory can be
used for graphene \cite{Samhedi} and we can write $$U_{sp^2}=\int
dxdy\, W_0[\bar{u}(x,y)]$$
by parametrizing the deformed surface as
$\bar{u}(x,y) = [u(x,y), v(x,y), w(x,y)]$. 
The elastic energy density $W_0$ can be divided into stretching and bending
contributions, $W_0 = W_0^S+W_0^B$.
The stretching energy density can be written as \cite{Atalaya_thesis}
\begin{equation} \label{eq:Eq_W_stretching}
W_0^{S} = \frac{Eh}{2(1-\nu^2)} [ E_{xx}^2 + E_{yy}^2 + 2\nu E_{xx}E_{yy} + 2(1-\nu)E_{xy}^2],
\end{equation}
where the components of the Green strain tensor $E_{ij}$ are
\begin{eqnarray}
E_{xx} &=& u_{x}+(u_{x}^2 + v_{x}^2+w_{x}^2)/2,\nonumber \\
E_{xy} &=& (u_{y} + v_{x} + u_{x}u_{y} + v_{x}v_{y} + w_{x}w_{y})/2,\nonumber \\
E_{yy} &=& v_y + (u_{y}^2 + v_{y}^2 + w_{y}^2)/2. \nonumber
\end{eqnarray} 
Here the subscripts on $u,v,w$ denote differentiation, i.e.
$u_x=\partial u/\partial x$ etc., and the coefficient $Eh$ represents
the Young modulus multiplied by the thickness of thin plate theory
\cite{Book_Drozdov,Book_Landau}. \emph{ It is important to note that
  in our theory $Eh$ is a single parameter and we do not consider any
  thickness in our formulation}. The Poisson ratio is denoted by
$\nu$. Both $Eh$ and $\nu$ are related to the Lam\'{e} parameters
$\lambda$ and $\mu$ as $Eh=4 \mu (\lambda + \mu)/(\lambda + 2\mu)$ and
$\nu = {\lambda}/(\lambda + 2\mu)$. The Lam\'{e} parameters are
determined from $\alpha$ and $\beta$ as $6\mu = {\sqrt{3}}({\alpha} +
8\beta)$ and $6\lambda = {\sqrt{3}}({\alpha} - 4{\beta})$ (for details, see
\cite{Atalaya_thesis}).  Using values from Ref.\cite{Graphene_Lobo} ($\alpha =
155.9$ J/m$^2$ and $\beta=25.5$ J/m$^2$), gives $\mu = 103.89$ J/m$^2$
and $\lambda = 15.55$ J/m$^2$. The continuum theory is in an excellent agreement
with molecular dynamics simulations of graphene \cite{Samhedi}.

The bending energy density can be approximated as 
\[
W_0^{B} = \frac{\kappa}{2}(2a_0 H)^2,
\]
where $\kappa = \sqrt{3}a_0^2 \gamma/2 \approx 0.8$ eV is the bending ridigity and $H = (w_{xx} + w_{yy})/2$ is the local mean curvature. 

In clamped graphene-based NEMS applications (e.g: mechanical
resonators), we find that the stretching energy is much greater than
the bending energy.  In thin plate theory, 
the bending regime (linear theory) is valid only for out-of-plane
deflections less than the plate thickness \cite{Book_Landau}.
Following this argument, for graphene the
linear regime is almost nonexistent and the nonlinear stretching
regime is dominant. We have determined this to be the case for deflections
in excess of 0.5 $\mathrm{\AA}$. Thus, in the following we neglect the 
bending contribution.

For a given applied external body force $\bar{F}_0$, the equilibrium
shape can be obtained by direct minimization of the free energy
functional
$${\mathcal F}[\bar{u}(x,y)]=\int dxdy W_0-\int dx dy \rho_0
{m_c}^{-1} \bar{F}_0 \cdot \bar{u},$$
where $\rho_0$ is the mass
density and $m_c$ is the carbon mass. Alternatively, the equilibrium
shape may be found by solving for the dynamics of the system including
dissipation. We will here work with the latter approach.  The dynamic
equation in the stretching regime, where Eq.
(\ref{eq:Eq_W_stretching}) is used as the elastic energy density, can
be obtained from standard variational principles. The corresponding
equation of motion, with an added phenomenological damping term $c
\dot{\bar{u}}(x,y)$, is
\begin{equation} \label{eq:Eq_dynamics}
\ddot{\bar{u}}(x,y) +c\dot{\bar{u}}(x,y)= \rho_0^{-1}{\mathcal D} \hat{P}[\bar{u}(x,y)] + {m_c}^{-1} \bar{F}_0(x,y,t),
\end{equation}
where $\hat{P}$ is the Piola stress tensor. In cartesian coordinates ($xy$), the Piola stress tensor is 
\begin{eqnarray} 
P_{xx} &=& (1+u_x)[(\lambda+2\mu)E_{xx}+\lambda E_{yy}]+2 \mu u_y E_{xy},\nonumber \\
P_{xy} &=& 2\mu(1+v_y)E_{xy} + v_x[(\lambda+2\mu)E_{xx}+\lambda E_{yy}],\nonumber \\
P_{xz} &=& \lambda w_x (E_{xx}+E_{yy})+2\mu w_x E_{xx} + 2 \mu w_y E_{xy},\nonumber \\
P_{yx} &=& 2\mu (1+u_x)E_{xy} + u_y [(\lambda+2\mu)E_{yy}+\lambda E_{xx}],\nonumber \\
P_{yy} &=& (1+v_y)[(\lambda+2\mu)E_{yy}+\lambda E_{xx}] + 2 \mu v_x E_{xy},\nonumber \\
P_{yz} &=& \lambda w_y (E_{xx}+E_{yy}) + 2 \mu w_y E_{yy} +  2 \mu w_x E_{xy},
\label{eq:Eq_P_full}
\end{eqnarray}
and the linear differential operator ${\mathcal D}$ acts on $\hat P$ as 
$${\mathcal D}\hat{P}=\sum_{\chi=x,y,z}(\partial_x P_{x \chi}+\partial_y
P_{y \chi})\hat{\chi}.$$
Dirichlet boundary conditions can be imposed through $\bar{u} = \bar{
 u}_0(t)$ and Neuman boundary conditions through $ \bar{n}_0 \cdot
\hat{P} = \bar{P}_0(t)$, where $\bar{u}_0(t)$ and $\bar{P}_0(t)$ are
specified on the boundary.

We refer to  Eqs. \eqref{eq:Eq_dynamics} and \eqref{eq:Eq_P_full}
as the full nonlinear system or general elasticity equations. Since
the general expresions for the Piola stress tensor are quite
cumbersome and make the equations difficult to solve. The full equations
can be approximated by a simpler set of equations where we
neglect the second order contributions of the in-plane displacements
$u$ and $v$. This yields the following expressions for the
components of the strain tensor
\begin{eqnarray}
E_{xx} &\approx& u_{x} + w_{x}^2/2,\nonumber \\
E_{xy} &\approx& (u_{y} + v_{x} + w_{x}w_{y})/2,\nonumber \\
E_{yy} &\approx& v_y + w_{y}^2/2, 
\label{eq:Eq_S_vK}
\end{eqnarray} 
and the Piola tensor 
\begin{eqnarray} 
P_{xx} &\approx& (\lambda+2\mu)E_{xx}+\lambda E_{yy},\nonumber \\
P_{xy} &\approx& 2\mu E_{xy},\nonumber \\
P_{xz} &\approx& \lambda w_x (E_{xx}+E_{yy}) + 2\mu (w_x E_{xx} + w_y E_{xy}),\nonumber \\
P_{yx} &\approx& 2\mu E_{xy},\nonumber \\
P_{yy} &\approx& \lambda E_{xx} + (\lambda+2\mu)E_{yy},\nonumber \\
P_{yz} &\approx& \lambda w_y (E_{xx}+E_{yy}) + 2 \mu (w_y E_{yy} + w_x E_{xy}). 
\label{eq:Eq_P_vK}
\end{eqnarray}
The resulting equations of motion for $u$, $v$ and $w$ are known in
thin plate theory \cite{Book_Drozdov,Book_Landau} as the nonlinear von
Karman equations. It is worth mentioning that we have arrived at these
equations without treating the graphene as a thin plate with some
thickness. 

We may further simplify the above expressions by completely
removing the in-plane displacements. The resulting nonlinear equation
for $w(x,y,t)$ is given by
\begin{equation} \label{eq:Eq_CET_simp}
\ddot{w}(x,y,t) + c\dot{w}(x,y,t) - \rho_0^{-1}\sum_{\chi = x ,y} \partial_{\chi} (w_{\chi} T_{\chi})  = \frac{F_{0z}}{m_c},
\end{equation}
where
\begin{eqnarray}
T_x &=& (\lambda + 2\mu)\delta_x + \lambda \delta_y + (\lambda/2 + \mu)(w_{x}^2 + w_{y}^2), \nonumber \\
T_y &=& (\lambda + 2\mu)\delta_y + \lambda \delta_x + (\lambda/2 + \mu)(w_{x}^2 + w_{y}^2). \nonumber
\end{eqnarray}
Here we have introduced the constants $\delta_x$ and $\delta_y$
representing initial strains in the $x$ and $y$ directions,
respectively. Such strains may appear during the
manufacturing process of graphene resonators. The functions $T_x$ and
$T_y$ are tensions in the $x$ and $y$
directions induced by stretching of the graphene. We refer to Eq.
\eqref{eq:Eq_CET_simp} as the out-of-plane approximation.

If one mode is expected to dominate the out-of-plane deformations, the
out-of-plane approximation may be projected onto this mode to
obtain an {ordinary} rather than a {partial} differential
equation. If the dominant mode is the fundamental, we can write
for a square sheet ($-a \le x,y, \le a$)
\[ w(x,y,t) = w(t)\cos(\frac{\pi x}{2a})\cos(\frac{\pi y}{2a}).
\]
By using this Ansatz, we obtain a Duffing equation for the amplitude of the fundamental mode,
\begin{equation}
\ddot{w}(t) + c\dot{w}(t) + \omega_0^2 w(t) + \frac{5 \pi^4 (\lambda + 2\mu)}{128 a^4 \rho_0} {w}^3(t) = \frac{F(t)}{m_c},
\label{eq:Eq_Duffing}
\end{equation}
where 
\[
F(t) = \frac{1}{a^2} \iint F(x,y,t)\cos(\frac{\pi x}{2a})\cos(\frac{\pi y}{2a}) dxdy
\]
is the overlap of the driving force with the shape of the fundamental mode, and the resonant frequency $\omega_0$ is given by 
\[
\omega_0= \frac{\pi}{a}\sqrt{\frac{(\lambda + \mu)(\delta + \delta^2/2)}{\rho_0}}
\]
and we have assumed $\delta_x = \delta_y =\delta$. Notice if several
modes are included in the Ansatz, we obtain a set of coupled
Duffing-type equations for the mode amplitudes.

\noindent {\bf Numerical results: Statics.}  
We now proceed to investigate the accuracies of the different approximations 
--- full nonlinear system (\ref{eq:Eq_dynamics}-\ref{eq:Eq_P_full}),
von Karman approximation (\ref{eq:Eq_S_vK}-\ref{eq:Eq_P_vK}),
out-of-plane approximation (\ref{eq:Eq_CET_simp}), and the Duffing 
equation (\ref{eq:Eq_Duffing}).  For illustration, we consider in
the simulations a clamped square graphene sheet of side $2a=1$ $\mu$m, (Dirichlet boundary condition with
$\bar{u}_0(t) = 0 $ on all edges) subject to uniform pressure $F_{dc}\hat{z}/A$ where
$A=3\sqrt{3}a_0^2/4$ is the area per carbon atom in graphene.
In all computations we consider a pre-tensile graphene with
initial strains of $\delta_x = \delta_y = 0.5\%$. We solve
the Eq. \eqref{eq:Eq_dynamics} numerically by using the spectral method
\cite{Book_hpEM,Yosibash_Collocation_Method} with 81 global basis
functions. 

\begin{figure}[hbtp]
\centering
\includegraphics[height=9cm,angle=-90,clip]{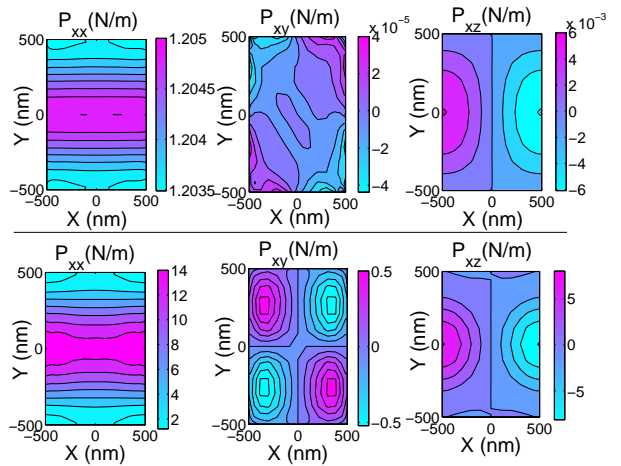}
\caption{Tension distributions for small deflections with $F_{dc} = 1$ fN per carbon atom (upper plots) 
and large deflections with $F_{dc} = 1$ pN (lower plots). Note the very different colour scales.}
\label{fig:fig_statics_1} 
\end{figure}

\begin{figure}[hbtp]
\centering
\vspace*{0.3cm}
\includegraphics[width=5cm,angle=270]{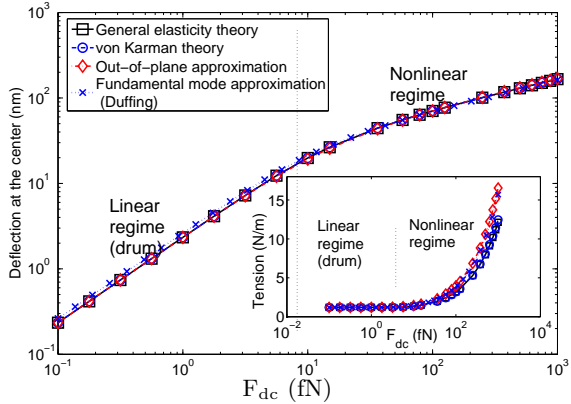}\\
\small{F$_{\rm dc}$ (fN)}\\
\caption{Elastostatics study for clamped pre-tensile graphene. The figure shows the vertical deflection at the center of the graphene sheet in response to a uniform force ranging from 0.1 fN to 1 pN per atom. Good agreement is obtained between the simplified models and the full nonlinear system. The inset shows the dependence between the tension on  the driving force.}\vspace*{-0.3cm}
\label{fig:fig_statics_2} 
\end{figure}

The tension in the graphene sheet varies depending on how much
stretching is imposed by the external forces. For
pre-tensile graphene, small deflections do not change the
built-in tension and the graphene-based resonator behaves as a
linear tensile elastic membrane \cite{Timoshenko}. However, when the deflections are large enough, the stress
 changes significantly from the initial value. To illustrate this, we present Figs.
\ref{fig:fig_statics_1} and \ref{fig:fig_statics_2}. The first figure
shows the stress distributions for two values of driving forces
$F_{dc}$ per carbon atom, $F_{dc} = 1$ fN and $F_{dc} = 1$ pN. For the
smaller force, the deflection at equilibrium is small and the tension
$P_{xx}$ is almost uniform and equal to the initial value 1.2 N/m. This 
agrees with the theoretical value of $\sigma = Eh/(1-\nu) \delta
\approx 1.2$ N/m, where $\delta = 0.5 \%$ is the initial strain of
the pre-tensile graphene. On the other hand,
for the larger force $F_{dc} = 1$ pN, the equilibrium built-in tension $P_{xx}$, as well as the
other components of the Piola stress tensor, is not
uniform and it varies from 2 N/m to 14 N/m. The inset in Fig.
\ref{fig:fig_statics_2} shows the dependence between the average value
of the tension $P_{xx}$ and the driving force $F_{dc}$. Note that for
small $F_{dc}$ the tension is almost constant and equal to 1.2 N/m,
but for large $F_{dc}$ the tension varies as $F_{dc}$ as $P_{xx} \sim
F_{dc}^{2/3}$.

Next, we present the results for the vertical deflection at the center
of the graphene sheet as function of the external force $F_{dc}$. We
find that for a pre-tensile graphene sheet the response is
linear for small values of $F_{dc}$. This is
due to an almost uniform tension, independent of
the amount of deflection, see Fig. \ref{fig:fig_statics_2}. On the
other hand, when the out-of-plane deflections are large, we see a
nonlinear response $w \sim F_{dc}^{1/3}$, where $w$ is the deflection
at the center of the graphene.

Finally, evaluations of the bending energy for the equilibrium shapes,
for a range of $F_{dc}$ between 0.1 fN and 1 pN, show that the
stretching energy is at least four orders of magnitude larger than the
bending energy. This validates the assumption that bending energy in
clamped graphene-based nanoresonators may be neglected. In addition,
we find that all the simplified mechanical descriptions show good
agreement with the general theory in the static case.

\noindent {\bf Numerical results: Dynamics.}  In this section, we study the
dynamic response of pre-tensile clamped graphene subject to small and
large deformations. For small deformations, when the tension is set by
the initial strain, the clamped graphene sheet behaves as a linear
tensile elastic membrane \cite{Timoshenko}. On the other hand, when the out-of-plane
deformations are large and tension depends on the amount of
deflection, a nonlinear response governs the dynamic behaviour. Here
we show that the dynamics exhibits Duffing-type behavior, e.g. we
observe the characteristic instability in the amplitude when we sweep
the driving frequency.

We divide the analysis into three parts. First, we consider a driving
force with the same spatial distribution as the fundamental mode. Here
we have only one mode present which makes this case easy to analyse.
In the second part we consider a uniform force where several modes
couple to the external force. Here we observe the presence of up to
three modes with similar Duffing-type response as the fundamental
mode. Finally, we study small oscillations around an equilibrium
determined by the dc-component of the exciting forces. We evaluate the
resonance frequencies as function of the $F_{dc}$ and the amplitude of
oscillation as a function of the ac-component of the driving force
$F_{ac}$.

\begin{figure}
\centering
{\includegraphics[width=8.2cm]{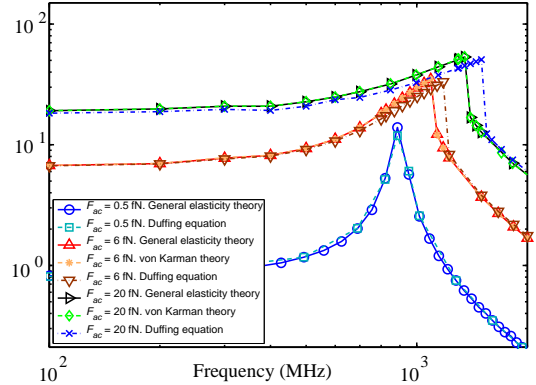}}
\caption{Amplitude of the fundamental mode for different driving forces strength. Observe the amplitude instability for large values of $F_{ac}$. All points were obtained starting from a configuration with the sheet at rest. If the dynamic state of the system is kept between frequency steps a hysteretic response obtains.}
\label{fig:fund_mode_comp} 
\end{figure}

For a driving force with the same spatial dependence as the fundamental mode,
\[ 
F(x,y,t) = F(t)\cos(\frac{\pi x}{2a})\cos(\frac{\pi y}{2a}),
\]
the out-of-plane and Duffing approximations are equivalent. We perform
numerical computations for small and large harmonic driving forces
($F(t) = F_{ac}\cos(\omega t)$).  The results are shown in Fig.
\ref{fig:fund_mode_comp}. The figure shows that for small driving
force $F_{ac} = 0.5$ fN, the response is that of a linear harmonic
resonator.  However, for the larger driving forces of $F_{ac} = 6$ fN
and 20 fN, we see a nonlinear response which manifests as an amplitude
instability at a critical, amplitude dependent frequency; e.g., at $f
= 1100$ MHz for $F_{ac} = 6$ fN. The Duffing equation predicts this
instability at a slightly higher frequency. This is because the
tension is slightly reduced when the in-plane displacements are
included (general elasticity theory or von Karman approximation), see
Fig. \ref{fig:fig_statics_2}. There is an excellent agreement between
the general elasticity theory and the von Karman approximation.

\begin{figure}
\centering
\includegraphics[width=5cm,angle=270]{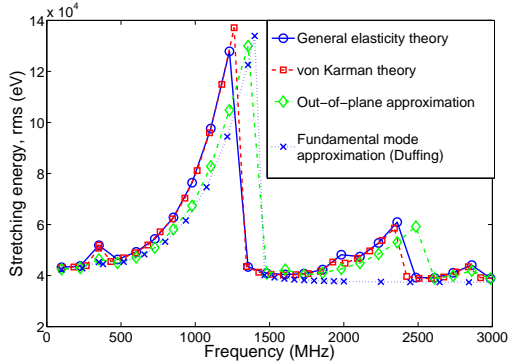}
\caption{Stretching energy vs. driving frequency for uniform driving force $F_{ac} = 10$ fN.}
\label{fig:stret_enrg_uni_comp} 
\end{figure}

Next, we study the case of a uniform driving force. Here several modes
couple to the driving forces. The fundamental mode has the greatest
overlap followed by the modes
\[\cos(\frac{3\pi x}{2a})\cos(\frac{\pi y}{2a}), \cos(\frac{\pi x}{2a})\cos(\frac{3 \pi y}{2a}),\]
and 
\[\cos(\frac{3\pi x}{2a})\cos(\frac{3\pi y}{2a}).\]
These modes are found by plotting the stretching energy against the
driving frequency. This is shown in Fig.
\ref{fig:stret_enrg_uni_comp}. The figure shows three peaks
corresponding to the different modes: the peak at 1200 MHz belongs to
the fundamental mode, and the peaks at 2360 MHz and 2840 MHz
correspond to the other modes mentioned above. Notice that the first
peak at 350 MHz does not correspond to any mode but to a $\frac{1}{3}$
sub-harmonic of the fundamental. The presence of this sub-harmonic is
due to the cubic nonlinear term in Eq. \eqref{eq:Eq_Duffing}.

Finally, we study small oscillations around an equilibrium shape
determined by a dc-part of the driving force $F_{dc} +
F_{ac}\cos(\omega t)$, typical for a NEMS resonator. We assume a
spatially uniform driving force. We find that the resonance frequency
for small oscillations increases as we increase the stretching or
deflection at the equilibrium position. This may be used to tune the
frequency of the oscillator by changing the dc-bias as $f_{res}
\propto F_{dc}^{1/3}$.

\begin{figure}
\centering
\includegraphics[width=5cm,angle=270]{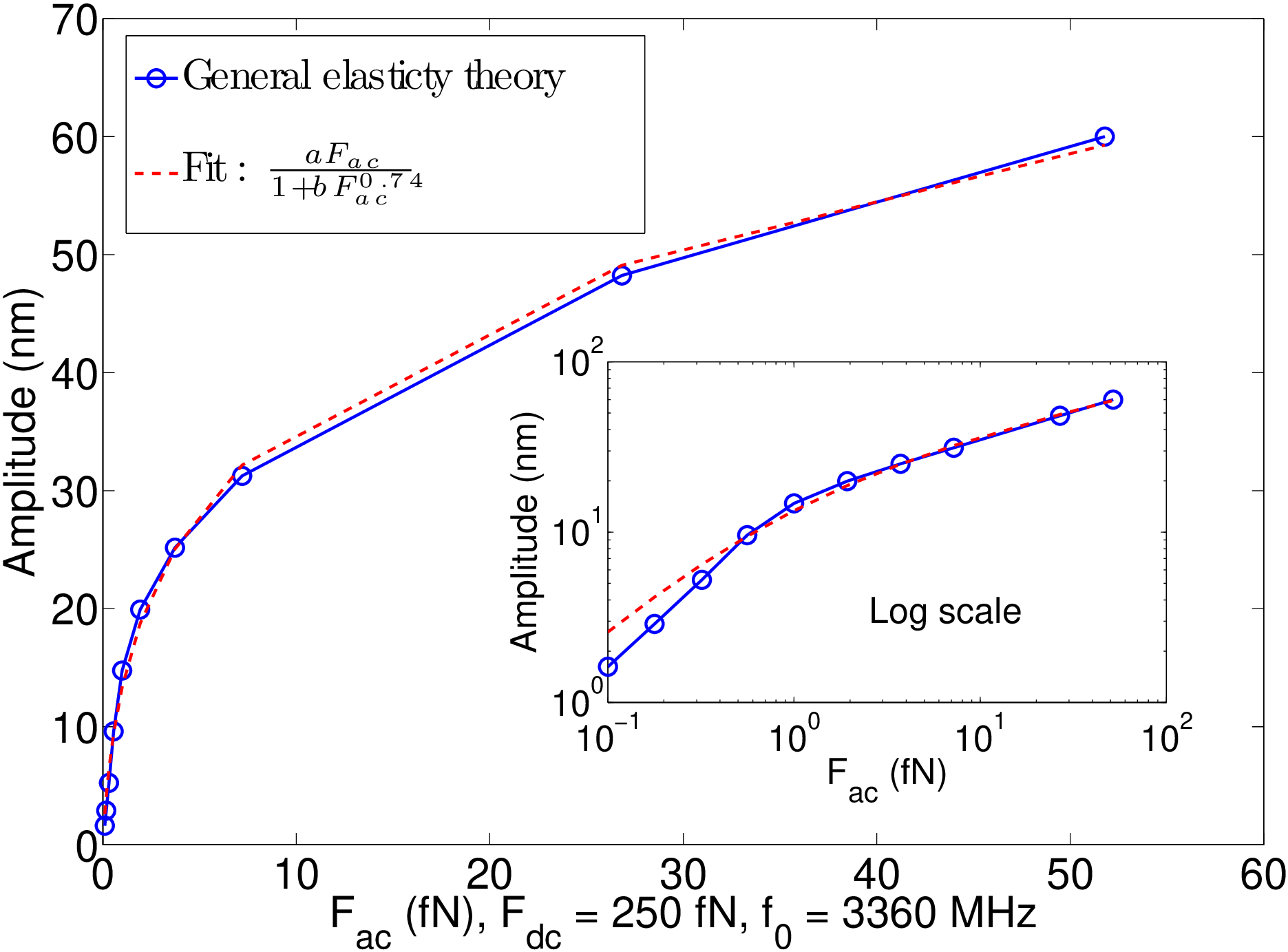}
\caption{Amplitude of small oscillations vs. $F_{ac}$.}
\label{fig:amp_fac} 
\end{figure}

Next, we study the dependence between the amplitude of the
oscillations as function of the driving ac-force $F_{ac}$. We find
that there is a linear response for small $F_{ac}$ and then it tends
to follow the law $F_{ac}^{1/3}$. This is shown in Fig.
\ref{fig:amp_fac}. Here the driving frequency is the resonance
frequency for $F_{dc}=250$ fN, which is 3360MHZ.

{\bf Conclusions.} We have derived and analyzed a nonlinear finite
elasticity theory for graphene resonators, both for elastostatics and
elastodynamics problems. Moreover, we have studied how this general
elasticity theory can be simplified to more easily solvable equations.
In particular, the out-of-plane approximation, Eq.
\eqref{eq:Eq_CET_simp}, gives good agreement with the general
elasticity theory while maintaining the advantage of being
computationally efficient. We have also shown that the dynamic
response of clamped graphene resembles that of coupled Duffing-type
resonators.

{\bf Acknowledgements.} 
We acknowledge fruitful discussions with Herre van der Zant, Menno Poot, 
Kaveh Samadikhah and Ener Salinas. We are grateful to the Swedish 
Foundation for Strategic Research and the Swedish Research Council for 
their financial support.

\end{document}